\documentstyle[aps,prl]{revtex}
\begin{document}
\input epsf.sty
\twocolumn[\hsize\textwidth\columnwidth\hsize\csname %
@twocolumnfalse\endcsname
\draft
\widetext
%%%%%%%% prl (above) %%%%%%%%%%%%%%%%%%

\title{Structural Critical Scattering Study of Mg-Doped CuGeO$_3$}

\author{Y. J. Wang, V. Kiryukhin, R. J. Birgeneau}
\address{Department of Physics, Massachusetts Institute of Technology,
Cambridge, MA 02139}
\author{T. Masuda, I. Tsukada, K. Uchinokura}
\address{Department of Applied Physics, The University of Tokyo,
7-3-1 Hongo, Bunkyo-ku, Tokyo 113-8656, Japan}

\date{\today}
\maketitle

\begin{abstract}

We report a synchrotron x-ray scattering study of 
the diluted spin-Peierls (SP) material Cu$_{1-x}$Mg$_x$GeO$_3$. 
We find that for $x>0$ the temperature T$_m$ at which
the spin gap is established 
is significantly higher than the
temperature T$_s$ at which the SP dimerization attains long-range order.
The latter is observed only for $x<x_c\sim$0.021,
while for $x>x_c$ the SP correlation length quickly decreases
with increasing $x$. We argue that
impurity-induced competing interchain interactions
play a central role in these phenomena.

\end{abstract}


%%%%%% prl format (below) %%%%%%%%%%%%%%
\phantom{.}
]
\narrowtext
%%%%%%%%% prl (above) %%%%%%%%%%%%%%%%%%

Low-dimensional quantum spin systems exhibit a variety of intriguing  
and often counter-intuitive properties.
A prominent example of such a material is the spin-Peierls (SP) system \cite{SP}
which consists of an array of one-dimensional (1D) antiferromagnetic
spin-chains with S=$1\over 2$ on a deformable 3D lattice. Below the
spin-Peierls transition temperature, T$_{SP}$, the spin-chains dimerize
and a gap opens in the magnetic excitation spectrum. The discovery of an 
inorganic SP compound CuGeO$_3$ (Ref. \cite{Hase})
made possible a systematic study of impurity
effects on SP systems \cite{Hase2}. In CuGeO$_3$, 
Zn$^{2+}$, Mg$^{2+}$ (S=0), and Ni$^{2+}$ (S=1) 
can be substituted for Cu$^{2+}$ (S=$1\over 2$)
thus directly affecting the
spin-chains \cite{Luss,Masuda}. 
A variety of unexpected phenomena that cannot be
explained by a simple dilution effect were quickly discovered in
impurity-doped CuGeO$_3$ \cite{Luss}-\cite{Martin}. At present,
a comprehensive theoretical description of the T-$x$ phase
diagram of impurity-doped
CuGeO$_3$ still remains to be
established. 

Despite the extensive amount of work on doped CuGeO$_3$,
the phase diagram of this system is still not fully established and remains
controversial, especially at large doping. Notably, there is a disagreement
between the results obtained by different experimental techniques. In 
Cu$_{1-x}$Zn$_x$GeO$_3$, for instance, the SP transition appears to 
vanish above  
$x\sim$0.02 as evidenced by
magnetic susceptibility measurements \cite{Hase2}, but appreciable
lattice dimerization is still observed at $x$=0.047 in neutron diffraction
experiments \cite{Martin}. 
Diffraction measurements also show that at large $x$    the SP
state does not attain long range order (LRO) in the case of Zn and Ni doping
\cite{Martin,Kir},  
but no systematic data are available.
Recently, Masuda {\it et al.} \cite{Masuda}
reported magnetic susceptibility measurements
indicating that 
the Mg-doped material undergoes a first order transition 
as a function of $x$ at $x_c\sim$0.023.
According to Ref. \cite{Masuda}, 
the SP transition abruptly disappears at $x_c$.  
This basic scenario has been confirmed by Nakao {\it et al.} \cite{Nakao}
using neutron diffraction techniques; however, they argue that 
$x_c$=0.027$\pm$0.001 and that broadened SP dimerization peaks 
are still present for $x>x_c$ in Cu$_{1-x}$Mg$_x$GeO$_3$. 
It is evident that a
reliable determination of the SP transition temperature T$_{SP}$
is of crucial importance. 
A second order phase transition temperature is typically
defined as the temperature at which the correlation length diverges and
LRO is achieved in the system. Therefore,
to resolve the current controversy on the phase diagram of doped CuGeO$_3$ 
and to elucidate the underlying physics,
application of an experimental technique 
such as high resolution synchrotron x-ray scattering
which probes with high precision any possible long-range structural
order is needed.

In this paper, we report a synchrotron x-ray scattering study
of the temperature-doping phase diagram of Cu$_{1-x}$Mg$_x$GeO$_3$. 
We find the surprising result that except for the
pure compound, 
the temperature T$_s$ at which the SP dimerization attains LRO is significantly
lower than
the temperature T$_m$ at which both 
short-range order SP dimerization
diffraction peaks appear {\it and} a local spin gap is established \cite{gap}.
The opening of the magnetic gap is deduced from the susceptibility
measurements of Ref. \cite{Masuda}. 
We argue that the structural properties of doped CuGeO$_3$
are determined by impurity-induced competing interchain
interactions and, possibly, random fields, and therefore are
similar to the properties of other systems with competing interactions
and/or fields, such
as Spin Glasses and Random Field Ising Model (RFIM) compounds.
These random competing interchain interactions naturally produce significant
differences in the local and long-range ordering temperatures.
The SP LRO is observed only for $x<x_c$ with $x_c\lesssim$0.021.
In fact, in this paper we define $x_c$ using this criterion; see also Ref. 
\cite{Masuda2}.
For $x>x_c$, the SP correlation length quickly decreases and the SP peaks
weaken with increasing $x$. Thus, in agreement with 
the hypothesis in Ref. \cite{Masuda}, the SP
transition is absent for $x>x_c$. 

The experiment was carried out at MIT-IBM beamline X20A at the 
National Synchrotron Light Source. The 8.5 keV
x-ray beam was focused by a mirror,
monochromatized by a pair of Ge (111) crystals, scattered from the sample,
and analyzed by a Si (111) analyzer. High
quality \cite{Masuda2} Cu$_{1-x}$Mg$_x$GeO$_3$
single crystals (0.01$^\circ$-0.02$^\circ$ mosaic width)
from the same batches as those studied in \cite{Masuda} 
with $x$ varying from zero to 0.035 were used. Carefully cleaved samples
were placed inside a Be can filled with helium heat-exchange gas and
mounted on the cold finger of a 4K closed cycle cryostat.   
The experiment was carried out around
the (1.5, 1, 1.5) SP dimerization peak position with the
(H K H) zone in the scattering plane. 
The highest intrinsic
correlation length  measurable in our experiment
was about 0.5$\mu$m (5000$\rm\AA$); 
we considered that a sample possessed LRO if the correlation
length exceeded this value since significantly larger lengths are 
essentially macroscopic.

%%==============================================================================
\begin{figure}
\centerline{\epsfxsize=2.9in\epsfbox{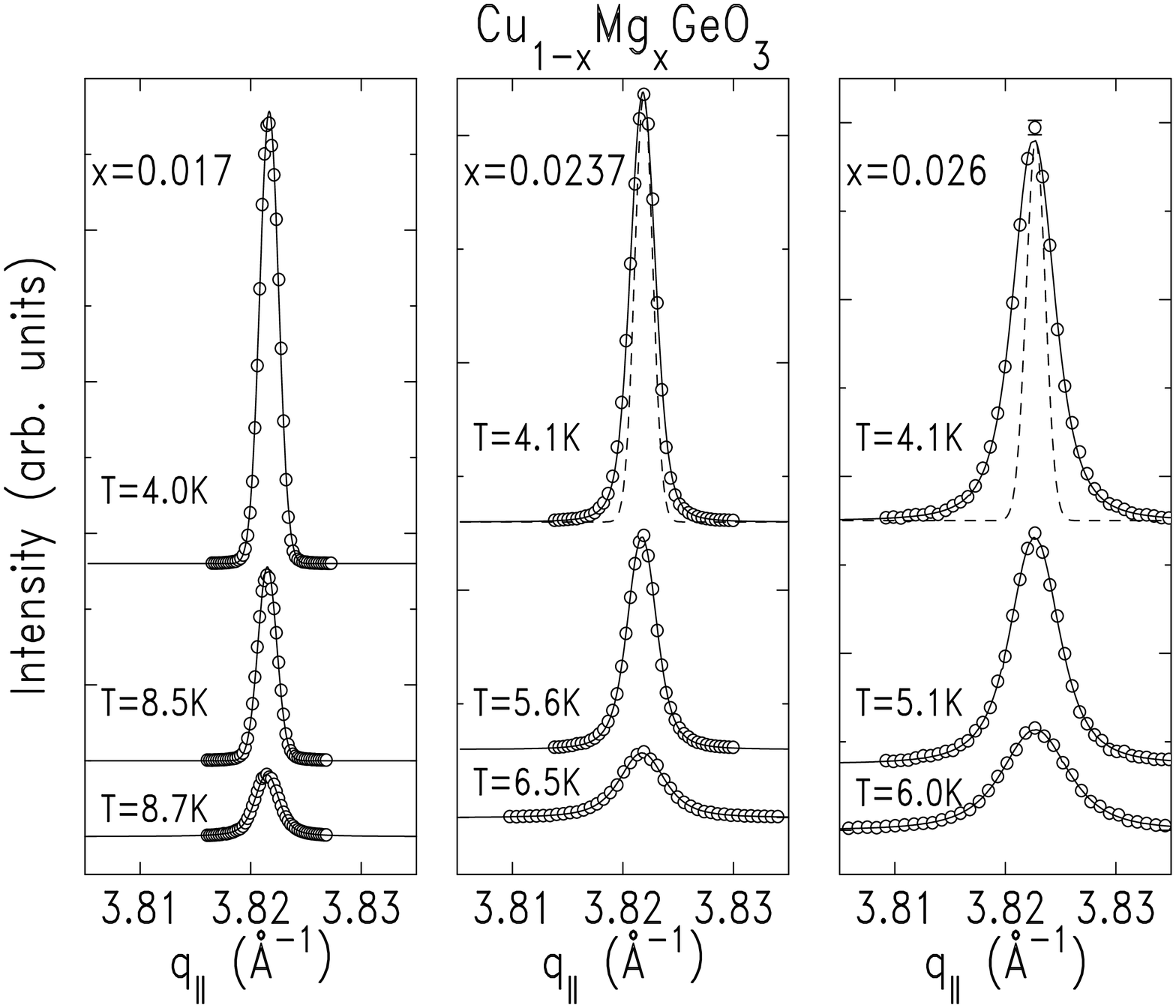}}
\vskip 5mm
\caption{Representative longitudinal scans at the (1.5, 1, 1.5) SP dimerization
peak position.
The solid lines are the results of fits as discussed in the text.
The instrumental resolution function is shown as a dashed line. }
\label{fig1}
\end{figure}
%%==============================================================================

A representative set of longitudinal (parallel to the scattering vector) 
x-ray scans at the (1.5, 1, 1.5) 
SP dimerization peak position for $x$=0.017, 0.0237, and 0.026 samples at
various temperatures is shown in Fig. 1. 
The low-temperature SP peak intensity in the $x$=0.017 sample was 
4$\times 10^4$ cts/sec with a background of 10 cts/sec.
The low-temperature SP peak width is resolution-limited in the $x$=0.017
sample, indicating that the SP state possesses LRO.
On the other hand, the SP peaks in the $x$=0.0237 and
$x$=0.026 samples are broadened at all experimentally accessible
temperatures, and therefore only
short-range order is present.
To extract the intrinsic correlation
length, the data for all samples were fitted to a two-dimensional convolution
(longitudinal and transverse directions) of the experimentally measured
Gaussian resolution function with the
intrinsic cross section (Lorentzian line-shapes taken to various powers were
attempted). The intensity was automatically integrated
over the third direction due to our experimental setup. A presumed 3D 
Lorentzian-squared intrinsic line-shape resulted
in the best fits; they are shown as the solid lines in Fig. 1. The
Lorentzian squared cross section is characteristic of 3D systems with 
quenched randomness \cite{Yosh},  
thus indicating that randomness plays a
central role in the phase behavior in the doped SP samples.          

%%==============================================================================
\begin{figure}
\centerline{\epsfxsize=2.7in\epsfbox{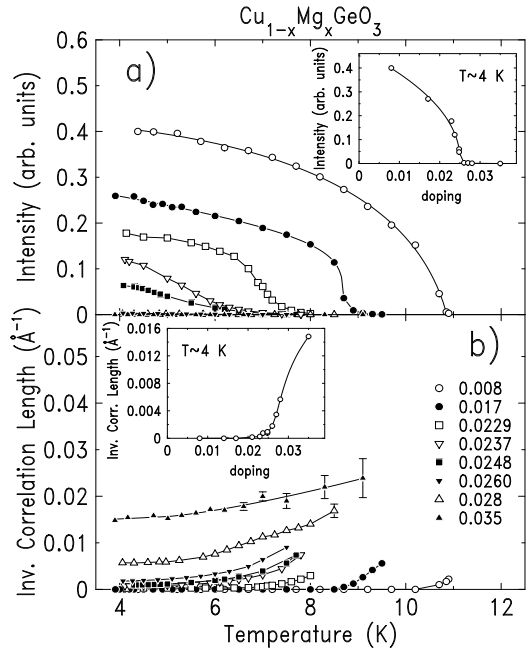}}
\vskip 5mm
\caption{The (1.5, 1, 1.5) SP peak intensity (a) and the corresponding
inverse correlation length (b) as functions of temperature for various Mg-doping
concentrations. Insets show the impurity-concentration
dependence of the low-temperature
SP peak intensity (a), and the corresponding inverse correlation length (b).
The solid lines are guides to the eye.}
\label{fig2}
\end{figure}
%%==============================================================================

Fig. 2(a) shows the temperature dependence of the (1.5, 1, 1.5)
dimerization peak intensity
for various doped samples. The data are normalized to the corresponding
(1, 0, 1) Bragg peak intensity. In contrast to the case of pure CuGeO$_3$,
the transitions are noticeably rounded with the rounding increasing with
increasing $x$ \cite{Kir2}. Such a substantial rounding cannot be
explained by local concentration gradients since the doping is too low to
create any correlated effect. The inset in Fig. 2(a) shows the maximum SP
peak intensity as a function of doping $x$. The peak intensity abruptly
decreases around $x\sim$0.025, which is roughly the doping level at which the SP
dip in susceptibility measurements is reported to disappear \cite{Masuda}. 
Fig 2(b) shows the temperature
dependence of the longitudinal inverse correlation length. The 
low-temperature inverse correlation length as a function of $x$ is shown in the
inset. The system attains LRO only for $x<x_c\lesssim$0.021. While substantial
intensity 
is still observed at the SP peak position for $x>x_c$,  
the system has only short range SP order.
Therefore, the phase transition as a function of $x$ reported by 
Masuda {\it et al.} \cite{Masuda} is characterized
by the loss of SP LRO at the critical doping $x_c$.  
As noted previously,
a similar conclusion has been reached by Nakao {\it et al.} albeit with
$x_c\sim$0.027; we believe that their results are consistent with ours
and that the difference in the deduced values for $x_c$ 
originates primarily in the much
higher resolution of our synchrotron x-ray study compared with the neutron 
diffraction experiments.
As discussed in Ref. \cite{Masuda2}, other anomalies observed by Nakao
{\it et al.} \cite{Nakao} for $x\sim$0.027 are connected with the
termination of the pseudo two-phase coexistence region of the N\'eel
state.

A central observation of this work is illustrated in Fig. 3 which presents 
data for the $x$=0.021 sample. This sample attains LRO at low temperatures.
Fig. 3 shows the temperature dependences of the
SP peak intensity, the intensity at the wing of the SP peak,
and the corresponding longitudinal intrinsic peak width.
A phase transition
temperature is properly 
defined as the temperature at which the correlation
length of the order parameter fluctuations diverges. It is evident from the
data of Fig. 3 that the temperature at which the LRO is achieved in this sample
is noticeably lower than both the onset temperature at which significant
intensity is observable at the SP peak position and the temperature
at which the wing intensity shows a peak. Moreover, magnetic 
susceptibility and neutron diffraction measurements 
give an even higher apparent SP
transition temperature T$_{SP}\sim$11K \cite{Masuda,Nakao}. 
The discrepancy between the neutron
and x-ray diffraction measurements can again 
be understood by taking into account
the moderate momentum-space resolution of the former technique. 
If, as in the neutron experiment,
we plot the SP peak intensity integrated over the corresponding broad 
momentum-space resolution
volume \cite{Kir2}, we obtain an apparent
transition temperature which is much higher than the one
evinced by  the data of Fig. 3. 
Qualitatively, this can be seen from the data of Fig. 3 by taking into
account the typical resolution of a neutron scattering experiment 
(0.003-0.005 $\rm\AA^{-1}$) and the fact that the integrated intensity
is roughly equal to the product of the peak intensity and the two in-plane
peak widths. 
Thus, the putative SP transition temperature found in the
diffraction experiments from the SP peak intensity {\it depends on the 
instrumental resolution}. However, the transition temperature can be 
consistently determined as the temperature at which the system attains LRO
as deduced from high-resolution x-ray measurements. Susceptibility
measurements determine the temperature T$_m$ at which a {\it local} spin gap
appears \cite{gap}
and this may or may not be simply related to the temperature at which
true long-range structural order occurs \cite{Dagotto}.

Unlike pure CuGeO$_3$, doped samples exhibit anomalously large relaxation
times resulting in marked thermal hysteresis.                            
The hysteresis disappears if the data are taken sufficiently slowly. Typically,
a 30 to 60 min wait at each temperature is enough to insure the 
absence of measurable
hysteresis. The slow dynamics found in this system will be discussed
in much more detail elsewhere \cite{Kir2}. 

%%==============================================================================
\begin{figure}
\centerline{\epsfxsize=2.7in\epsfbox{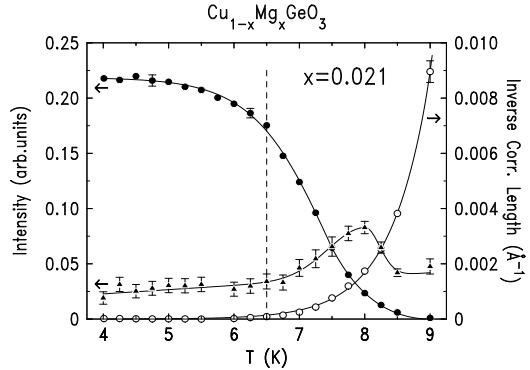}}
\vskip 5mm
\caption{The temperature dependences of the (1.5, 1, 1.5) SP peak intensity
(filled circles), the intensity at the wing ($\Delta q$=4.5$\times
10^{-3}\rm\AA^{-1}$) of the SP peak (filled triangles),
and the corresponding longitudinal
inverse correlation length (open circles) in
the $x$=0.021  sample.
The dashed line indicates the temperature at which, to the accuracy of
our experiment, the long-range order is achieved.
The solid lines are guides to the eye.}
\label{fig3}
\end{figure}
%%==============================================================================

We have carefully determined the \hfill impurity-concentration dependence
of the temperature T$_s$ at which the LRO is achieved. The experiments
were performed slowly enough to insure the absence of 
measurable thermal hysteresis.
The results, together with the susceptibility measurements 
of Ref. \cite{Masuda},
are presented in the revised phase diagram for Cu$_{1-x}$Mg$_x$GeO$_3$
shown in Fig. 4.
The error bars in Fig. 4 include possible beam heating effects
estimated in Ref. \cite{Kir2}.
These beam heating effects may also result in a small renormalization
of the temperature scale in Figs. 2 and 3. 
{\it Two} characteristic temperatures, both related to the SP transition 
but defined by the physics on different length-scales, are found
in the doped material. At the
higher temperature, T$_m$, the magnetic susceptibility
and other thermodynamic probes, such as specific heat exhibit an anomaly
reflecting the opening of a local
magnetic gap at T$_m$ \cite{gap}. 
While at high $x$ these probes give large uncertainties 
in the value for T$_m$,
{\it low resolution} diffraction probes indicate the appearance of the lattice
dimerization at about the same temperature, providing additional 
evidence for the local opening of the magnetic gap. As discussed above,
we find that this lattice dimerization around T$_m$ has a finite correlation
length. The LRO is attained at a lower temperature T$_s$, at which no
significant anomaly in the magnetic susceptibility  is found. 
Thus, the following
chain of events occurs with decreasing temperature. First, 
at T=T$_m$ local
dimerization occurs and the magnetic gap opens up in finite-sized domains
which presumably extend between individual Mg$^{2+}$ dopants. 
As the temperature decreases and the 3D interactions favoring the SP state
become relatively more important, the individual 
domains between the random 
Mg$^{2+}$ dopants on different chains begin to correlate
with each other over large distances, and 3D LRO is eventually achieved 
at T=T$_s$. Since this process involves only local rearrangement
in the phase of the dimerization, no significant 
anomaly in thermodynamic variables is expected at T$_s$.   
We should emphasize that we cannot rigorously
exclude the scenario in which for $x<x_c$
the low-temperature
SP correlation length 
saturates at some finite value which is 
larger than our resolution limit of 0.5$\mu$m.
In this case, T$_s$ would be characterized as a long distance 
cross-over temperature rather than a true second order
phase transition temperature.

The rich and complex properties exhibited by doped CuGeO$_3$ clearly cannot
be explained by simple dilution effects alone. In fact, our data suggest that
the SP ``transition''
in the doped material cannot be described as a conventional equilibrium phase
transition. The severe rounding of the transition, the 3D 
Lorentzian-squared peak shapes, and the anomalously slow dynamics
found in Cu$_{1-x}$Mg$_x$GeO$_3$
closely resemble analogous phenomena found in 
disordered systems with competing fields or
interactions, such as the 
RFIM compounds 
\cite{Birgeneau}, and Spin Glasses \cite{Mydosh}. 
We believe that 
the resemblance of doped CuGeO$_3$ to 
systems with competing interactions or fields is not coincidental. 
As first argued by Harris {\it et al.} \cite{Harris}, the
SP system can be mapped onto an effective 3D Ising model, in which the two dimer
configurations possible in a given chain are associated with the up and
down states of the Ising spins. 
The interchain interaction keeps the
spin-chain dimerizations
in phase. When a Cu atom is replaced by Mg in an isolated chain, 
it is energetically
favorable for the Ising pseudospin direction to change sign
across the impurity since the Mg
ion makes one of the original dimer configurations energetically unfavorable.   
Concomitantly, because the interchain interaction favors all pseudospin-up 
(or all pseudospin-down)
configuration, introduction of impurities creates frustration in the 3D
system. 
This is analogous to the situation in 3D Ising systems with mixed
ferromagnetic and antiferromagnetic bonds although the appropriate 
pseudo-spin model needed to describe Cu$_{1-x}$Mg$_x$GeO$_3$
must be more elaborate since
one must consider explicitly the differences in behavior of even and odd
number Cu$^{2+}$ 1D spin segments as well as higher order random field effects.
We will discuss this model much more
extensively in Ref. \cite{Kir2}. At the minimum, it is clear that
competing interactions are intrinsic to the doped CuGeO$_3$ system
and must be included in any realistic model describing this 
compound \cite{Most}.

%In summary, using high-resolution synchrotron x-ray scattering techniques, 
%we have
%established a revised phase diagram of the doped SP compound 
%Cu$_{1-x}$Mg$_x$GeO$_3$. 
%In the doped material, a spin gap opens and local SP dimerization appears
%at a temperature noticeably higher than the temperature at
%which the SP dimerization attains 
%long-range order. We present a possible
%microscopic scenario elucidating this behavior.   
%The SP LRO ($\xi >$0.5$\mu$m) 
%is found only for $x<x_c\sim$2.1\%. For $x>x_c$, the low-temperature
%SP correlation length quickly decreases and the SP peaks weaken with increasing
%$x$. These results elucidate the microscopic nature of the transition
%recently reported by Masuda {\it et al}. \cite{Masuda}. 
%We believe that 
%impurity-induced competing interactions play an essential 
%role in these novel phenomena.  

%%==============================================================================
\begin{figure}
\centerline{\epsfxsize=2.8in\epsfbox{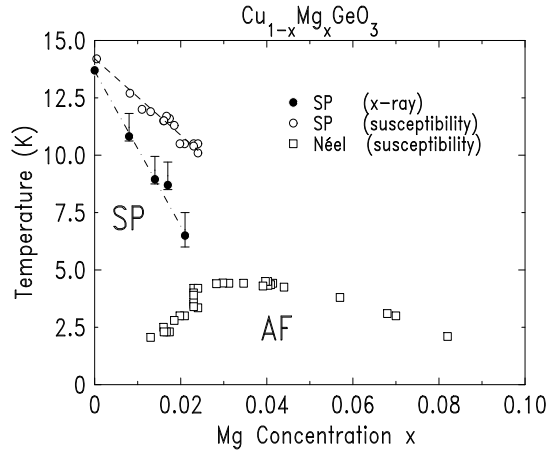}}
\vskip 5mm
\caption{The revised temperature-doping phase diagram of
Cu$_{1-x}$Mg$_x$GeO$_3$.
The filled circles indicate the temperature T$_s$ at which the SP long-range
order is achieved as determined in our experiment.
The open circles show the
SP transition temperature T$_m$ as inferred from magnetic susceptibility
measurements,
and the squares show the N\'eel temperature [5].}
\label{fig4}
\end{figure}
%%==============================================================================

We are grateful to G. Shirane, Y. Fujii, J. P. Hill, L. D. Gibbs, and A. Aharony
for important discussions. We thank S. LaMarra for assistance 
with the experiments. This work was
supported by the NSF under Grant No. DMR97-04532.
This work was also supported by
NEDO 
International Joint Research Grant, and by Grant-in-Aid for COE Research
``SCP Coupled System'' from
the Ministry of Education, Science, Sports, and Culture of
Japan.

%%%%%%%%%%%%%%%%%%%%%%%%%%%%%%%%%%%%%%%%%%%%%%%%%%%%%%%%%%%%%%%%%%%%%%%%


\begin{references}
%%%%%%%%%%%%%%%%%%%%%%%%%%%%%%%%%%%%%%%%%%%%%%%%%%%%%%%%%%%%%%%%%%%%%%%%

\bibitem{SP}
For a review, see J. W. Bray {\it et al.} in {\it Extended Linear Chain 
Compounds}, ed. J. C. Miller (Plenum, NY, 1982)

\bibitem{Hase}
M. Hase, {\it et al.}, Phys. Rev. Lett. {\bf 70}, 3651 (1993)

\bibitem{Hase2}
M. Hase, {\it et al.}, Phys. Rev. Lett. {\bf 71}, 4059 (1993)

\bibitem{Luss} 
J.-G. Lussier, {\it at al.}, J. Phys. Cond. Mat. {\bf 7}, L325 (1995);
M. Hase, {\it et al.}, Physica B {\bf 215}, 164 (1995); S. B. Oseroff,
{\it et al.}, Phys. Rev. Lett. {\bf 74}, 1450 (1995); J. P. Schoeffel,
{\it et al.}, Phys. Rev. B {\bf 53}, 14971 (1996)

\bibitem{Masuda}
T. Masuda, {\it et al.}, Phys. Rev. Lett. {\bf 80}, 4566 (1998)

\bibitem{Manabe}
K. Manabe, {\it et al.}, Phys. Rev. B {\bf 58}, R575 (1998)

\bibitem{Kir}
V. Kiryukhin, {\it et al.}, Phys. Rev. B {\bf 54}, 7269 (1996)

\bibitem{Martin}
Y. Sasago, {\it et al.}, Phys. Rev. B {\bf 54}, R6835 (1996);
M. C. Martin, {\it et al., ibid.}  {\bf 56}, 3173 (1997)

\bibitem{Nakao}
H. Nakao, {\it et al.}, preprint cond-mat/9811324

\bibitem{gap} in fact, inelastic neutron scattering measurements are 
needed to establish opening of the magnetic gap unambiguously

\bibitem{Masuda2} 
T. Masuda, {\it et al.}, preprint cond-mat/9906130

\bibitem{Yosh}
R. A. Cowley, {\it et al.}, Z. Phys. B {\bf 58}, 15 (1984);
U. Nowak, K. D. Usadel, Phys. Rev. B {\bf 46}, 8329 (1992)

\bibitem{Kir2}
V. Kiryukhin, {\it et al.}, unpublished

\bibitem{Dagotto} see also the discussion in G. B. Martins, {\it et al.},
Phys. Rev. B {\bf 54}, 16032 (1996)

\bibitem{Birgeneau}
R. J. Birgeneau, J. Mag. Mag. Mat. {\bf 177-181}, 1, (1998)

\bibitem{Mydosh} see, {\it eg.}, J. A. Mydosh, {\it Spin Glasses: 
an Experimental Introduction}, (Taylor \& Francis, London, 1993)

\bibitem{Harris} Q. J. Harris, {\it et al.}, Phys. Rev. B {\bf 52},
15420 (1995)

\bibitem{Most}
M. Mostovoy and D. Khomskii, Z. Phys. B {\bf 103}, 209 (1997)


\end{references}
\end{document}